\documentclass[prl,twocolumn]{revtex4}

\usepackage[dvips]{graphicx}%
\usepackage{bm,color}
\usepackage{amsmath,amssymb}
\newcommand{\braket}[2]{\langle #1 | #2 \rangle}
\newcommand{\ketbra}[2]{\ket{#1}\bra{#2}} 
\newcommand{\ket}[1]{\left |  #1 \right \rangle}
\newcommand{\bra}[1]{ \left \langle #1  \right |}

\def \tr{{\textrm {Tr}}}

\begin{document}
\title{Fundamental Quantum Limits for Practical Devices}

\author{Ryo Namiki}\affiliation{Department of Physics, Graduate School of Science, Kyoto University, Kyoto 606-8502, Japan}

\date{\today}
\begin{abstract} 
We present experimentally testable quantum limitations on the phase-insensitive linear amplification and phase conjugation  with respect to the transformation of a Gaussian distributed set of coherent states following the footing to assess the success of continuous-variable quantum teleportation and quantum memory devices.  The results enable us to compare the real device with the quantum limited device via feasible input of coherent states.  
\end{abstract}

\maketitle



An important role of theoretical physics is to derive the fundamental limitation on the performance of physical devices for manipulating the states of physical system. The controllability of the physical states over the existence of the quantum noise and its connections to quantum measurement are central objective in wide area of quantum physics 
 \cite{Cle10}. 
An elementary operation for signal processing is amplification and its quantum limitation is generally determined based on the canonical commutation relation \cite{amp}.
A pertinent approach is optimal cloning of quantum states so as to address the limitation on amplifying quantum information \cite{rmp-clone,Cerf00,Cer05}.
Those limitations are thought to be in the reach of experiments 
 \cite{amp2,Cle10,Sab07}.

Any physical process is described by a completely positive trace-preserving map referred to as quantum channel \cite{NC00,CV-RMP}. We often use the average fidelity as a figure of merit to estimate the performance of the process in quantum information science. The problems to find the quantum limit  phase-insensitive linear amplifier and to optimize the cloning map for coherent states are equivalent when the figure of merit is the trace norm \cite{Gut06} or the joint fidelity \cite{Cer05,namiki07}. In the case of the most familiar amplification limit,  the figure of merit is the ratio of the signal-to-noise ratios of the input and output fields \cite{amp}.
   Besides the amplification, an interesting quantum-state manipulation 
    is the phase conjugation \cite{Cer01a,Bus03}. It corresponds to the universal not gate for qubit states \cite{Buz99} and to a transposition map for finite and infinite dimensional states \cite{Bus03}. 
   
In contrast to those active signal processing, elementary devices for quantum communication and computation are designed to transfer quantum states in a rather passive  manner. Actually, ideal quantum memory or quantum teleportation process is an identical map, which retrieves the input states  without disturbance, and main step of quantum computation is to perform the  unitary operations, which implies reversible transformation of quantum states.  
Toward the realization of useful quantum devices, a cornerstone is to prove the quantum coherence of the process by beating the classical limit achieved by the classical measure-and-prepare (MP) schemes \cite{Bra00,Ham05,namiki07,namiki08,Has10,Takano08}. 
The MP scheme is an entanglement breaking channel \cite{16} and surpassing the classical limit fidelity is a proof of entanglement.
It is known that the optimal fidelity of the phase conjugation 
 can be achieved by a classical device \cite{Buz99,Bus03,Cer01a} and that the Gaussian phase-conjugation (time-reversal) map belongs to the entanglement breaking channel \cite{Hol08}. 

 To experimentally test the performance of the quantum device, an accessible input  state is the coherent state. 
 It is theoretically simple to determine the classical (or other physical) limitation assuming a uniform set of input states because the figure of merit has a covariant property \cite{Cer05} and the group theoretical treatment is useful \cite{Chi05}. However,
 neither testing the input-output relation for any coherent state nor assuming the displacement covariant property for the real device is feasible. In practical, available power of the input field is limited and the linearity of the real device is maintained only on a limited range of the input variable. In the case of optical or atomic  continuous-variable quantum information processing \cite{CV-RMP}, the amplitude of the  input coherent states has to be much smaller than the total photon number of the so-called local oscillator fields.
 To avoid the problem, a Gaussian distribution has been employed to observe the performance on a flat distribution of the coherent-state amplitude, thereby one can determine the device performance by using coherent states with a feasible amount of phase-space displacement \cite{Bra00,Ham05,namiki07}. The classical limit fidelity was initially determined for essentially identical process such as quantum teleportation and quantum memory \cite{Bra00,Ham05}, and generalized for a class of the non-unitary processes by considering a transformation task to take the effect of loss and amplification into account \cite{namiki07}. The classical limit was also determined for a class of multi-mode gates \cite{Takano08}. It is worth noting that the classical capacity for bosonic quantum channels has been derived under the  energy constraint \cite{Gio04}.  %

  Remarkably,  the quantum limitations on the amplification    \cite{amp,Gut06,Cer05,namiki07} and phase-conjugation \cite{Cer01a,Bus03} are presented based on the uniform distribution and covariant property. 
  In order to give a solid foundation as an experimental science, it is  crucial to address the quantum limitations under experimentally testable frameworks. Noting that the well-known quantum teleportation \cite{Furusawa98} and quantum memory \cite{Jul04} protocols serve as  amplifiers via the gain control mechanism, it is natural to work with the fidelity-based figure of merit. %

In this Letter, we consider the quantum limits of the phase-insensitive linear amplification and phase conjugation in terms of the average fidelity with respect to the Gaussian distributed set of coherent states.  We derive a tight quantum limit fidelity for the phase-insensitive amplification task and show that this fundamental limit is achieved by  the known Gaussian amplifier. We also derive a tight quantum limit fidelity for the phase conjugation task and show that this limit is achieved by a classical MP device. %

In what follows the state vector with the Greek letter ``$\alpha $'' denotes the coherent state and the state vector with the Roman letter ``$n$'' denotes the number state, e.g., we write the coherent state in the number basis as $\ket{ \alpha}= e^{-|\alpha |^2 /2} \sum_{n=0}^\infty \alpha ^n \ket{ n} /\sqrt{n!}$. When we work on the state with two modes, we call the first system $A$ and the second system $B$. 

Let us define the average fidelity of the physical process $\mathcal E$ for transformation task on the coherent states $\{|\sqrt N \alpha\rangle \} \to \{\ket{\sqrt \eta \alpha }\} $ with $N, \eta > 0$ by 
\begin{eqnarray}
F_{N, \eta, \lambda} (\mathcal E) &:=&  \int p_\lambda( \alpha )\bra{\sqrt \eta \alpha }\mathcal E \Big(|\sqrt N \alpha \rangle \langle \sqrt N \alpha| \Big) \ket{\sqrt\eta \alpha} d^2 \alpha  \nonumber \\ \label{eq1} 
\end{eqnarray}
where  the prior distribution of a symmetric Gaussian function with the inverse width of $\lambda >0 $ is given by \begin{eqnarray}
p_\lambda( \alpha ) :=  \frac{\lambda }{\pi} \exp (- \lambda |\alpha |^2 ).\label{eq2}\end{eqnarray} This distribution describes the uniform distribution in the limit $\lambda \to 0$. 
The fidelity represents the average probability that the input state $ | \sqrt N \alpha  \rangle $ is exactly transformed into the corresponding target state $ | \sqrt \eta  \alpha \rangle  $ by the process $\mathcal E$. When $\eta / N \ge 1 $, the transformation task implies  the amplification of the coherent-state amplitude with the gain factor $\eta / N $.
When $\eta = N =1$, the task is referred to as the unit-gain task and the fidelity estimates how well the input coherent state is retrieved at the output port. 
When $\eta / N < 1 $ the transformation suggests amplitude dumping. This is the case for practical transmission and storage processes, and the loss of fidelity can be seen as a deviation from the ideal lossy channel. When $N$ and $\eta $ are positive integers the task may be called $N$-to-$\eta$ cloning where the fidelity implies how well the transformation from $N$ copies $\ket{\alpha} ^{\otimes N}$ to $\eta$ copies $\ket{\alpha} ^{\otimes \eta}$ can be achieved. The \textit{quantum limit fidelity} is defined as an upper limit of the average fidelity $F ( \mathcal E)$ achieved by the completely positive trace-preserving map $\mathcal E$. We call the limit is \textit{tight} if the fidelity limit is achieved by a completely positive trace-preserving map.  
Note that, from Eqs. (\ref{eq1}) and (\ref{eq2}), by changing the integral parameter we can verify the identity: \begin{eqnarray}F_{N, \eta, \lambda}   &=&  F_{\frac{N}{\eta },1, \frac{\lambda}{\eta }}, = F_{1, \frac{\eta }{N}, \frac{\lambda}{N} }.\label{imm}\end{eqnarray} 

 \textit{Quantum optimal phase-insensitive linear amplifier.---} 
Let us consider the amplification task $\{ |{  \alpha }\rangle \} \to \{ \ket{{\sqrt \eta }\alpha} \} $ with the gain $\eta > 1 $.  
 In the following we show that the fidelity $F_{1,\eta, \lambda}$ is bounded above by  
$\frac{1+ \lambda }{ \eta}$ for sufficiently small $\lambda$ and that this bound is achieved by the know Gaussian amplifier.  Note that the tight quantum limit fidelity of attenuation task  with $\eta \in [0,1]$ is unity \cite{namiki07}. 

 \textit{Proof.---}
Let us consider the following integration \cite{namiki-up}  
 with the parameters $s  \ge 0 $, $0 \le \kappa  \le 1 $ and $0\le  \xi < 1 $, 
\begin{eqnarray}
 J_{ \mathcal E}(s, \kappa ,\xi ) 
&:=& \int d^2 \alpha  p_s(\alpha ) \bra{  \alpha}_A \bra{\kappa \alpha^*}_B \mathcal\nonumber\\ & &  \mathcal E_A \otimes I_B \left( \ket{\psi_\xi}\bra{\psi_\xi} \right ) \ket{\kappa \alpha^*}_B
 \ket{ \alpha}_A  \label{start}
\end{eqnarray}
where  $\ket{\psi_\xi}= \sqrt{1-\xi ^2} \sum_{n=0}^\infty \xi^n\ket{n}\ket{n}$ is the two-mode squeezed state and  $I$ represents the identity process. The integration can be connected to the average fidelity by 
\begin{eqnarray}
 J_{ \mathcal E}(s,\kappa ,\xi )   &=& \frac{s(1-\xi^2)}{\lambda}
 F_{N,1, \lambda} (\mathcal E ) \label{jj}
\end{eqnarray}
where the parameters are supposed to satisfy the following relations 
\begin{eqnarray}
\lambda &=& s+ (1-\xi^2)\kappa ^2,   \label{lam}\\
\sqrt N &=& \kappa \xi. \label{n}
\end{eqnarray}
From the condition $s\ge 0 $ with Eqs. (\ref{lam}) and (\ref{n}),  we have  \begin{eqnarray}
\frac{\lambda}{1-\xi ^2} \le N+ \lambda . \label{scondition}
\end{eqnarray}

We proceed to consider the upper bound of $J_{\mathcal E}$ instead of the upper bound of the fidelity $F(\mathcal E)$. 
For any physical process with the complete positivity and trace-preserving condition, $\rho_{\mathcal E} := \mathcal E \otimes I ( \ketbra{\psi_\xi}{\psi_\xi})$ is a density operator. Then, the maximum of $J_{\mathcal E}$ with respect to the optimization of the process $ {\mathcal E}$ is no larger than the maximum achieved by the optimization of the density operator $\rho_{\mathcal E}$ over the set of the whole physical states. Thus we have   
\begin{eqnarray} \sup_{\mathcal E }  J_{ \mathcal E}(s,\kappa ,\xi)  &\le& \max_{\rho_{\mathcal E}} \tr \left[ \rho  M  \right] = \|  M \| \label{cp1}   \end{eqnarray}
where  we define
\begin{eqnarray}
M:= \int p_s(\alpha) \ket{ \alpha  }\bra{ \alpha  } \otimes \ket{\kappa  \alpha ^*  }\bra{\kappa \alpha ^* }  d^2\alpha    \nonumber
\end{eqnarray}
and   $\| \cdot \|:= \max_{\braket{u}{u}=1} \bra{ u} \cdot \ket{u} $ stands for the maximum eigenvalue. 

Since $M $ is a two-mode Gaussian state, its maximum eigenvalue is given from the symplectic eigenvalues of its covariance matrix \cite{Adess07}. 
Let us define the covariance matrix of a density operator on the two-mode field $  \rho $ 
\begin{eqnarray}
\gamma_\rho := \langle \hat R \hat R^t + (\hat R  \hat R^t)^t \rangle _{  \rho}  - 2 \langle \hat R \rangle  \langle \hat R^t  \rangle _{ \rho}  \nonumber 
\end{eqnarray} 
where $\hat R := (\hat x_A,\hat p_A,\hat x_B, \hat p_B)^t $ is the set of the quadrature operators of the mode  $A$ and  mode $B$  whose elements satisfy the canonical commutation relations $[\hat x_A, \hat p_A]=i $ and $[\hat x_B, \hat p_B]= i $. %
Then, the covariance matrix of the operator $M $ is calculated to be 
\begin{eqnarray}
\gamma_{M } = \openone_4 +  \frac{2}{s }\left(
  \begin{array}{cc}
     \openone_2    &  \kappa  Z   \\
      \kappa   Z &  \kappa ^2 \openone_2   \\
  \end{array}
\right),  \nonumber 
 \end{eqnarray}
where $\openone_4 :=\textrm{diag} (1,1,1,1) $, $\openone_2 := \textrm{diag}(1,1)$ and $ Z :=  \textrm{diag} (1,-1)$. 
 In order to diagonalize this matrix we define a matrix $U(r)$ corresponding  to the two-mode squeezing operator $\hat U_r := e^{-i( \hat x_A  \hat p_B +\hat x_B \hat p_A  ) r } = e^{( \hat a ^\dagger \hat b^\dagger -\hat a \hat b  ) r } $ through the transformation 
\begin{eqnarray}
\hat U^\dagger \hat R \hat  U &= &\left(
  \begin{array}{cc}
     \cosh r  \openone_2  & \sinh r   Z  \\
       \sinh r   Z & \cosh r \openone_2   \\
  \end{array}
\right) \hat R   
  =:  U(r) \hat R.   \nonumber 
\end{eqnarray}
When the squeezing parameter satisfies $\tanh{2 r} = 2\kappa /(1+ s +  \kappa  ^2)$ the covariance matrix is diagonalized  as
$U(-r) \gamma_{M } U^t (-r) = \textrm{diag} (\nu_+,\nu_+ ,\nu_-, \nu_-)$
where  the symplectic eigenvalues are determined to be
\begin{eqnarray}
\nu_\pm = \left[ \sqrt{(1+s+  \kappa ^2 )^2 -4\kappa ^2}\pm (1-\kappa ^2) \right] /s. \nonumber 
\end{eqnarray} 
Therefore, the diagonal form of $M $ is  the product of the thermal states $T(\bar n _+ )\otimes T(\bar n  _- ) $ with the mean photon numbers $\bar n_\pm = (\nu_\pm - 1) /2 $ where the   thermal state with the mean photon number $\bar n $ is defined by 
\begin{eqnarray}
 T( \bar n )  :=  {\frac{1 }{ 1+ \bar n   }} \sum_{n =0}^{\infty} \left( \frac{\bar n }{ 1+ \bar n  } \right)^{n }    |n  \rangle \langle n  |. \nonumber \label{thermal} 
\end{eqnarray} 
 This implies the following form of the maximum eigenvalue with the help of Eqs.  (\ref{lam}) and (\ref{n}):
 \begin{eqnarray}
\|M    \|  &=& 4/{[(\nu_+ +1)(\nu_- +1 )] } \nonumber \\ &=& \frac{2s}{ N + \lambda +1 +\sqrt{(N + \lambda +1)^2 -4 N / \xi^2  } } . \nonumber
\end{eqnarray}

Using this relation, Eqs. (\ref{jj}), (\ref{scondition}), and (\ref{cp1})  
we have \begin{eqnarray}
\sup_{\mathcal E} F_{N,1,\lambda } (\mathcal E) &\le& \frac{ \lambda }{s(1-\xi ^2 )}\| M  \|  \nonumber \\ 
&\le & \frac{ 2(N+ \lambda )}{ N+\lambda +1 +\sqrt{(N+\lambda-1) ^2}}\nonumber \\
&=& \left\{  \begin{array}{cc}      N+ \lambda  & \textrm{if  }(N+\lambda ) \le 1  , \\      1 & \textrm{if  }  (N+\lambda ) > 1    . \\  \end{array}\right.
\end{eqnarray}
By taking the replacement $(N, \lambda) \to (1/\eta, \lambda /\eta )$ and using the identity of Eq. (\ref{imm}), we obtain the upper bound of the fidelity for the amplification task, 
\begin{eqnarray}
\sup_{\mathcal E}  F_{1,\eta,\lambda} (\mathcal E) 
 &\le& \left\{  \begin{array}{cc}      \frac{1+ \lambda}{\eta }  & \textrm{if }  \eta  \ge  1 +\lambda   , \\      1 & \textrm{if  } \eta  <  1 +\lambda  . \\  \end{array}\right. \label{amplim}
\end{eqnarray}

Next we consider the attainability of this bound.
The Gaussian amplifier with the gain $g= \cosh ^2 r \ge 1$ is defined by ${\mathcal A}_g (\rho ):= \tr_B [ U_r\rho \otimes \ketbra{0}{0}_B  U_r^\dagger  ]$. It transforms the coherent state
as 
 ${\mathcal A} _g (\ketbra{\alpha }{\alpha }) 
=  \frac{1 }{\pi (g-1 ) }\int e^{-\frac{| \beta |^2 }{g-1 }} \ketbra{\sqrt g \alpha + \beta}{\sqrt g \alpha + \beta } d^2\beta$.  
This implies
\begin{eqnarray}
F_{1,\eta, \lambda}({ \mathcal A}_g) &=& \frac{\lambda}{\lambda g +| \sqrt g- \sqrt{ \eta  }|^2} \nonumber \\  
&=& \frac{\lambda}{(\lambda +1) (\sqrt{ g}  - \frac{\sqrt \eta}{ \lambda +1 }  )^2+ \frac{\lambda  \eta}{ \lambda +1 }  }
\le  \frac{1+ \lambda}{ \eta }, \nonumber 
\end{eqnarray}
where the equality is achieved when 
$g= \eta / (1+ \lambda )^2 \ge 1 $.
Therefore, the upper one of Ineqs. (\ref{amplim}) is saturated by the Gaussian phase-insensitive amplifier if the distribution is sufficiently flat so as to satisfy $\eta \ge (1+\lambda )^2 $.  \hfill$\blacksquare$  

In the limit of  $\lambda  \to 0$, our fidelity reproduces the quantum limit  for the case of the uniform distribution $F_o= 1/\eta $ 
 \cite{namiki07,Cer05}. As we can see, the fidelity value $(1+ \lambda )/\eta  $ always exceeds the uniform limit $F_o$, and thus a naive comparison of the experimental fidelity with $F_o$ gives an illegal result or an overestimation on how well the experimental device is approximating the quantum limited device. In contrast, our result includes the effect of the finite distribution $\lambda$, and enables a legitimate estimation toward the fundamental quantum limitation.

\textit{Optimal phase conjugator.---} 
Let us consider the phase-conjugation task $\{| \sqrt N \alpha \rangle \} \to \{\ket{ \alpha ^* }\}$ with $N>0$ and define the fidelity $F_{N,1,\lambda}^* (\mathcal E ):= \int d^2 \alpha p_\lambda (\alpha )\bra{\alpha^* }\mathcal E(| \sqrt N \alpha\rangle\langle {\sqrt N \alpha}|)\ket{\alpha^* } $. We can show that the optimal fidelity is  given by 
\begin{eqnarray}
\sup_{\mathcal E} F_{N,1,\lambda}^* (\mathcal E ) = \frac{N+\lambda }{N+ \lambda + 1  } , \label{pclimit}
\end{eqnarray}
 and is achieved by the classical MP scheme 
 \begin{eqnarray}
\mathcal E_{MP}^* (\rho ):= \frac{1}{\pi} \int \bra{\alpha }\rho \ket{ \alpha} \ket{\frac{\sqrt{N} \alpha ^*}{N +\lambda} }\bra{\frac{\sqrt{N}\alpha ^*  }{N +\lambda} } d^2 \alpha . \label{pceb}
\end{eqnarray}


\textit{Proof.---} We start by defining $J_{ \mathcal E}^*(s,\xi,\kappa )  := \int d^2 \alpha  p_s(\alpha ) \bra{  \alpha^*}_A \bra{\kappa \alpha^*}_B \mathcal E_A \otimes I_B \left( \ket{\psi_\xi}\bra{\psi_\xi} \right ) \ket{\kappa \alpha^*}_B \ket{ \alpha^*}_A $ similarly to Eq. (\ref{start}). Here, different from the previous case we assume a weaker constraint of $\kappa \ge 0 $. This suggests the phase-conjugation task with either attenuation or amplification.
  Similar to Eq. (\ref{jj}) we can confirm the following relation with the help of Eqs.  (\ref{lam}) and  (\ref{n}): 
\begin{eqnarray}
 J_{ \mathcal E}^*(s,\kappa ,\xi )   &=& \frac{s(1-\xi^2)}{\lambda}
 F_{N,1, \lambda}^* (\mathcal E ). \label{jj-2}
\end{eqnarray}
 An upper bound of $ J_{ \mathcal E}^*(s,\xi,\kappa ) $ is given by the optimization of the density operator $\rho = \mathcal E \otimes I (\ketbra{\psi_ \xi}{\psi_ \xi}) $ over the physically possible states, namely, we have \begin{eqnarray} \sup_{\mathcal E}  J_{ \mathcal E}^* (s,\kappa ,\xi ) &=& \max_{\rho  }  \tr [  \rho  M^* ] 
 \le    \|  M ^* \|  \label{ff} \end{eqnarray}
where we define
\begin{eqnarray}
 M  ^* = \int p_s(\alpha) \ket{ \alpha  }\bra{ \alpha  } \otimes \ket{\kappa  \alpha  }\bra{\kappa \alpha }   d^2\alpha . 
\end{eqnarray}
This operator is also a two-mode Gaussian state, and its covariance matrix is calculated to be 
\begin{eqnarray}
\gamma_{M^* } = \openone_4 +  \frac{2}{s }\left(
  \begin{array}{cc}
     \openone_2    &  \kappa    \openone_2   \\
      \kappa     \openone_2 &  \kappa ^2 \openone_2   \\
  \end{array}
\right).  \nonumber 
 \end{eqnarray}
This covariance matrix can be diagonalized by a beamsplitter transformation, and the symplectic eigenvalues are determined to be
$(\nu_+, \nu_-) = (1, 1+2(1+\kappa ^2 )/s )$.  Hence, we have \begin{eqnarray}
\| M^*  \| =4/ [(\nu_++1) (\nu_-+1) ]=  \frac{s }{s+1+ \kappa  ^2 }.  \label{ongm}
\end{eqnarray}
Equations (\ref{ff}) and (\ref{ongm}) lead to 
\begin{eqnarray}
 \sup_{\mathcal E }  J_{ \mathcal E}^*(s,\kappa ,\xi ) 
&\le & \frac{s }{s+1+ \kappa ^2 }. \nonumber 
\end{eqnarray}
Using this relation and Eqs.  (\ref{lam}), (\ref{n}), (\ref{scondition}),  
  and,  (\ref{jj-2}) we  obtain the upper bound of the fidelity for the phase-conjugation task 
\begin{eqnarray}
 \sup_{\mathcal E }  F_{N,1, \lambda }^* (\mathcal E ) &\le& \frac{\lambda}{ (1-\xi^2)} \frac{1}{N+ \lambda +1 }  \le    \frac{N+\lambda }{N+ \lambda + 1  } .  \nonumber  
\end{eqnarray} On the other hand, this bound is achieved by the MP scheme of Eq. (\ref{pceb}), i.e.,  
$F_{N,1, \lambda } (\mathcal E_{MP}^* )=  \frac{N+\lambda }{N+ \lambda + 1  }$ holds.   
 We thus obtain 
 Eq. (\ref{pclimit}). 
\hfill$\blacksquare$

The value of the optimal fidelity for the covariant approach \cite{Cer01a,Bus03} is reproduced when we set $N=1$ and take the limit $\lambda \to 0$. Our result shows that the optimality of the classical device for the phase-conjugation task occurs beyond the case of the uniform distribution.
The optimality of the classical device suggests the coincidence of the   quantum limit and classical limit. Such a coincidence also occurs when the target states are orthogonal to each other 
\cite{namiki08}.  Note that, when the optimization of the state $\rho_{\mathcal E}$ in Eq. (\ref{cp1}) is limited over the positive-partial-transpose states \cite{namiki-up}, the value of the optimal fidelity corresponds to the value of the optimal fidelity for the phase-conjugation task. Hence, for many of the tasks whose target states are given by the transpose of the input states, it is likely that the gap between the quantum limit and   classical limit disappears.


In conclusion, we have presented quantum limitations on the phase-insensitive linear amplification and phase conjugation in terms of the average fidelity by assuming transformation tasks on a Gaussian distributed set of coherent states. Thereby, experimental test can be done by using coherent states with a finite amount of phase-space displacement on the same footing as the success criterion for continuous-variable quantum teleportation and quantum memory. It was also shown that both of the fidelity limits can be achieved by the known Gaussian machines and that the known results for the case of the uniform distribution are safely reproduced. The present results give a solid foundation to experimentally observe how well the real device approximates the quantum limited device in a legitimate manner.

R.N. acknowledges support from JSPS.


\begin{thebibliography}{}  

\bibitem{Cle10} A.A. Clerk  \textit{et al.,} \rmp 82 1155 (2010).

\bibitem{amp}C. M. Caves, \prd\textbf{26,} 1817 (1982).

\bibitem{rmp-clone} V. Scarani, S. Iblisdir, N. Gisin, and A. Ac\'in, \rmp \textbf{77,} 1225 (2005).
\bibitem{Cerf00}  N.J. Cerf, A. Ipe, and X. Rottenberg, \prl \textbf{85,} 1754 (2000);   N.J. Cerf and S. Iblisdir, \pra \textbf{62,} 040301(R)  (2000). 
\bibitem{Cer05} N.J. Cerf \textit{et al.,} 
\prl \textbf{95,} 070501 (2005).




\bibitem{amp2}V. Josse \textit{et al.,} 
 \prl \textbf{96,} 163602 (2006); R.C. Pooser  \textit{et al.,}  \prl \textbf{103,} 010501 (2009).

 \bibitem{Sab07}  U.L. Andersen, V. Josse, and G. Leuchs
 \prl \textbf{94,} 240503 (2005); S. Koike \textit{et al.,}  \prl \textbf{96,} 060504 (2006); M. Sabuncu, U.L. Andersen, and G. Leuchs,
 \prl \textbf{98,} 170503 (2007).

\bibitem{NC00}M. A. Nielsen and I. L. Chuang, \textit{Quantum Computation and Quantum Information}, (Cambridge University Press, Cambridge, 2000);
\bibitem{CV-RMP} 
 N.J. Cerf, G. Leuchs, and E.S. Polzik (eds), \textit{Quantum Information with Continuous Variables of Atoms and Light}, (Imperial College Press, 2007). 
 
 

\bibitem{Gut06}M. Guta and K. Matsumoto, Phys. Rev. A 74, 032305 (2006).

\bibitem{namiki07} R. Namiki, M. Koashi, and N. Imoto, \prl\textbf{101,} 100502 (2008).

\bibitem{Cer01a} N. J. Cerf and S. Iblisdir, \pra \textbf{64,} 032307 (2001); 

\bibitem{Bus03}F. Buscemi, G.M. D'Ariano, P. Perinotti, and M.F. Sacchi, Phys. Lett. A 314, 374 (2003).
\bibitem{Buz99}V. Buzek, M. Hillery, and R. F. Werner, Phys. Rev. A 60, R2626 (1999).





\bibitem{Bra00}S. L. Braunstein, C.A. Fuchs, and J. Kimble, J. Mod. Opt \textbf{47,} 267 (2000).
\bibitem{Ham05} K. Hammerer, M.M. Wolf, E.S. Polzik, and J.I. Cirac, \prl \textbf{94,} 150503 (2005).

\bibitem{namiki08}R. Namiki, \pra 78, 032333 (2008).
\bibitem{Takano08} T. Takano, M. Fuyama, R. Namiki, and Y. Takahashi, \pra 78, 010307(R) (2008).
\bibitem{Has10} H. H\"aseler and N. L\"utkenhaus, \pra 81, 060306(R) (2010).







\bibitem{16}M. Horodecki, P. W. Shor, and M. B. Ruskai, 
Rev. Math. Phys. {\bf 15}, 629-641 (2003).

\bibitem{Hol08} A. S. Holevo, 
 Probl. Inf. Transm. 44,  3, (2008).

\bibitem{Chi05}G. Chiribella, G.M. D'Ariano, P. Perinotti, and N.J. Cerf, Phys. Rev. A 72, 042336 (2005).


\bibitem{Gio04} H. P. Yuen and M. Ozawa, Phys. Rev. Lett. 70, 363 (1993); Giovannetti \textit{et al.,} Phys. Rev. Lett. 92, 027902 (2004).

\bibitem{Furusawa98}A. Furusawa \textit{et al}., Science \textbf{282,} 706 (1998); S.L. Braunstein and H.J. Kimble, \prl{\bf 80}, 869 (1998).

\bibitem{Jul04} B. Julsgaard \textit{et al.,} 
\nat 432, 482 (2004). 

\bibitem{namiki-up} R. Namiki, unpublished. 

 

\bibitem{Adess07}G. Adesso and F. Illuminati, J. Phys. A 40, 7821 (2007).






\end{thebibliography}
\end{document}